\documentstyle[11pt]{article} 

\def \inbar{\vrule height1.5ex width.4pt depth0pt}
\def \xC{\relax\hbox{\kern.25em$\inbar\kern-.3em{\rm C}$}}
\def \xR{\relax{\rm I\kern-.18em R}}

\newcommand{\xbe}{\begin{equation}}
\newcommand{\be}{\begin{equation}}
\newcommand{\xee}{\end{equation}}
\newcommand{\ee}{\end{equation}}
\newcommand{\xbea}{\begin{eqnarray}}
\newcommand{\bea}{\begin{eqnarray}}
\newcommand{\xeea}{\end{eqnarray}}
\newcommand{\eea}{\end{eqnarray}}

\begin{document}

\title{Comment on the Possibility of a Geometric Constraint in the Schr\"odinger Quantum Mechanics}
\author{Ali Mostafazadeh\thanks{E-mail address: 
amostafazadeh@ku.edu.tr}\\ \\
Department of Mathematics, Ko\c{c} University,\\
Rumelifeneri Yolu, 80910 Istanbul, TURKEY}
\date{ }
\maketitle

\begin{abstract}
It is shown that the geometric constraint advocated in [R.\ S.\ Kaushal, Mod.\ 
Phys.\ Lett.\ A {\bf 15} (2000) 1391] is trivially satisfied. Therefore, such a
constraint does not exist. We also point out another flaw in Kaushal's paper.
\end{abstract}
\vspace{2mm}


\baselineskip=24pt

In Ref.~\cite{ka}, the author uses the polar decomposition $\psi(x)=
Nu(x)e^{iS(x)}$ of the solution of the one-dimensional time-independent 
Schr\"odinger equation
	\be
	\psi''(x)+k^2(x)\psi(x)=0,
	\label{sch-eq}
	\ee
with $k^2:=2m[E-V(x)]/\hbar^2$ to derive a so-called geometric constraint, 
namely 
	\be
	K={\rm constant},
	\label{const}
	\ee
where
	\be
	K=c^2(\psi/u)^2+(\psi' u-u'\psi)^2\;,
	\label{K}
	\ee
and $c$ is an arbitaray constant.

Eq.~(\ref{K}) is obtained using the Schr\"odinger equation written in terms of
$\psi$ (i.e., Eq.~(\ref{sch-eq})) and in terms of the polar variables $u$ and $S$,
namely 
	\bea
	u''+[k^2-S'^2]u&=&0\;,
	\label{e1}\\
	uS''+2u'S'&=&0\;.
	\label{e2}
	\eea
As pointed out in \cite{ka}, Eq.~(\ref{e2}) can be integrated to yield
	\be
	S'=c/u^2\;.
	\label{e3}
	\ee
Substituting this equation in (\ref{e1}), one finds the Milne equation
	\be
	u''(x)+k^2(x)u(x)=c^2/u^3(x)\;,
	\label{milne}
	\ee
and eliminating $k^2$ from Eqs.~(\ref{sch-eq}) and (\ref{milne}) and
integrating the result one arrives at (\ref{const}).

This part of the analysis of \cite{ka} is correct. However, it is not difficult
to show that the right hand side of ({\ref{K}) vanishes. In order to see this, we 
use the identity
	\be
	\frac{\psi}{u}=Ne^{iS}\;,
	\label{e4}
	\ee
and Eq.~(\ref{e3}) to compute
	\be
	\psi' u-u'\psi=u^2\frac{d}{dx}\left(\frac{\psi}{u}\right)
	=iNu^2S'e^{iS}=icNe^{iS}.
	\label{e5}
	\ee
Now substituting Eqs.~(\ref{e4}) and (\ref{e5}), in the right hand side of
Eq.~(\ref{K}), we find that $K$ vanishes identically. Therefore, the
condition~(\ref{const}) is equivalent to the trivial identity: $0={\rm 
constant}$.

We wish to conclude this note by pointing out that the same analysis may be
used to simplify Eq.~(16) of Ref.~\cite{ka} to ${\cal K}=c_1^2(u/\psi)^2=
c_1^2e^{-2iS}/N^2$. Therefore, the claim that ${\cal K}$ is constant for the
Milne  equation, 
	\be
	\psi''(x)+k^2(x)\psi(x)=c_1^2/\psi^3(x),
	\label{e6}
	\ee
seems to indicate that this equation does not have a solution with a variable 
phase $S$. This claim can be easily rejected by constructing a solution of Eq.~(\ref{e6})
of the form $\psi(x)=e^{iS(x)}$ where $S(x)$ is any real and nonconstant solution of the
Mathieu equation
	\be
	S''(x)+c_1^2\sin[4S(x)]=0\;,
	\label{e7}
	\ee
and $k^2(x)$ is given by 
	\[k^2(x)=S^{'2}(x)+c_1^2\cos[4S(x)]\;.\]
We believe that there is a flaw in the derivation of Eq.~(16) of Ref.~\cite{ka}. It is 
not difficult to see that in the polar representation of the Milne equation~(\ref{e6}), one 
cannot easily decouple the equations for $u$ and $S$. Therefore, contrary to the claim made
in \cite{ka} a similar analysis does not seem to lead to Eq.~(16) or a corrected version 
of it.

\end{document}